\documentclass[12pt,epsf]{article}
\usepackage{graphicx}
\usepackage[dvipsnames]{color}
\usepackage{epsfig}
\usepackage{amsmath}
\usepackage{amssymb}
\usepackage{amsfonts}
\usepackage{subfigure}
\usepackage{slashbox}
\begin{document}

\def\a{\alpha}
\def\b{\beta}
\def\c{\varepsilon}
\def\d{\delta}
\def\e{\epsilon}
\def\f{\phi}
\def\g{\gamma}
\def\h{\theta}
\def\k{\kappa}
\def\l{\lambda}
\def\m{\mu}
\def\n{\nu}
\def\p{\psi}
\def\q{\partial}
\def\r{\rho}
\def\s{\sigma}
\def\t{\tau}
\def\u{\upsilon}
\def\v{\varphi}
\def\w{\omega}
\def\x{\xi}
\def\y{\eta}
\def\z{\zeta}
\def\D{\Delta}
\def\G{\Gamma}
\def\H{\Theta}
\def\L{\Lambda}
\def\F{\Phi}
\def\P{\Psi}
\def\S{\Sigma}

\def\o{\over}
\def\beq{\begin{eqnarray}}
\def\eeq{\end{eqnarray}}
\newcommand{\lsim}{\raisebox{0.6mm}{$\, <$} \hspace{-4.4mm}\raisebox{-1.5mm}{\em$\sim\,$}}
\newcommand{\gsim}{\raisebox{0.6mm}{$\, >$} \hspace{-3.0mm}\raisebox{-1.5mm}{\em $\sim\,$}}
\newcommand{\vev}[1]{ \left\langle {#1} \right\rangle }
\newcommand{\bra}[1]{ \langle {#1} | }
\newcommand{\ket}[1]{ | {#1} \rangle }
\newcommand{\EV}{ {\rm eV} }
\newcommand{\MeV}{ {\rm MeV} }
\newcommand{\GeV}{ {\rm GeV} }
\newcommand{\TeV}{ {\rm TeV} }
\def\diag{\mathop{\rm diag}\nolimits}
\def\Spin{\mathop{\rm Spin}}
\def\SO{\mathop{\rm SO}}
\def\O{\mathop{\rm O}}
\def\SU{\mathop{\rm SU}}
\def\U{\mathop{\rm U}}
\def\Sp{\mathop{\rm Sp}}
\def\SL{\mathop{\rm SL}}
\def\tr{\mathop{\rm tr}}

\def\IJMP{Int.~J.~Mod.~Phys. }
\def\MPL{Mod.~Phys.~Lett. }
\def\NP{Nucl.~Phys. }
\def\PL{Phys.~Lett. }
\def\PR{Phys.~Rev. }
\def\PRL{Phys.~Rev.~Lett. }
\def\PTP{Prog.~Theor.~Phys. }
\def\ZP{Z.~Phys. }

\def\Z{\mathcal{Z}}
\def\W{\Omega}
\def\alert#1{\textcolor{red}{#1}}

\def\stau1{\tilde{\tau}_{1}}
\def\bino{\tilde{B}}
\def\n1{\tilde{\chi}^{0}_{1}}


\baselineskip 0.7cm

\begin{titlepage}

\begin{flushright}
\end{flushright}

\vskip 1.35cm
\begin{center}
{\large \bf
Relaxing the Higgs mass bound in singlet extensions\\ of the MSSM \\
}
\vskip 1.2cm
Kazunori Nakayama$^{1}$, Norimi Yokozaki$^{1}$ and Kazuya Yonekura$^{1,2}$
\vskip 0.4cm

{\it
$^1$Department of Physics, University of Tokyo, 
Tokyo 113-0033,
Japan\\

$^2$ Institute for the Physics and Mathematics of 
the Universe (IPMU),\\  
University of Tokyo, Chiba 277-8568, Japan\\ 

}

\vskip 1.5cm

\abstract{
We show that the upper bound on the lightest Higgs mass in the MSSM is relaxed by introducing a singlet which couples to the Higgs fields, even at a large $\tan\beta$ region, preferable for explaining the muon anomalous magnetic moment.
In the models of a singlet extension, it is known that the upper bound is relaxed by a tree-level contribution, 
especially at small tan$\beta$ region.
For large tan$\beta$, however, the requirement for the perturbativity on the singlet-Higgs coupling 
up to the GUT scale prevents the lightest Higgs from obtaining a large tree-level mass.
We construct an explicit UV complete model which allows large singlet-Higgs coupling at low energy
without disturbing the perturbativity. The UV completion can be applied for any singlet extension of the MSSM. 
Moreover, we point out that the radiative correction from the singlet-Higgs coupling becomes dominant, and
 the lightest Higgs mass can be easily as heavy as 130 GeV 
if this coupling is large enough even for large tan$\beta$.
}
\end{center}
\end{titlepage}

\setcounter{page}{2}

\section{Introduction}

In the minimal supersymmetric (SUSY) standard model (MSSM)~\cite{Martin:1997ns},
the lightest Higgs boson mass cannot be heavier than the $Z$-boson mass at tree level.
By taking account of the radiative correction, the upper bound on the Higgs mass
is relaxed~\cite{Okada:1990vk}
and it can satisfy the LEP bound : $m_h \gtrsim 114.4$\,GeV.
However, it is extremely difficult to raise the lightest Higgs mass up to $\sim 130$\,GeV
in the framework of MSSM with TeV scale SUSY. Recent results from ATLAS and CMS collaborations may indicate
 the relatively heavy Standard Model like Higgs boson.

The bound on the Higgs mass is relaxed in a slight modification to the MSSM.
One of the simplest possibilities is to add a singlet field ($S$) to the MSSM, which couples to the Higgs fields 
in the superpotential as\footnote{Another possibility is to add vector-like matters to MSSM. In the case that the vector-like matters couple to the up-type Higgs with $\mathcal{O}(1)$ Yukawa coupling, the radiative corrections from the vector-like matters can raise the lightest Higgs mass significantly~\cite{Moroi:1991mg,Babu:2004xg}.}
\begin{equation}
	W = \lambda S H_u H_d,   \label{SHH}
\end{equation}
where $\lambda$ is a coupling constant.
This type of extension is motivated from the viewpoint of $\mu$-problem,
since if the singlet scalar $S$ obtains a vacuum expectation value (VEV) of $\langle S\rangle \sim$ TeV,
it explains the origin of $\mu$-parameter.
There are variety of ways to impose symmetries such that bare $\mu$-term is prohibited while
the coupling (\ref{SHH}) is allowed.
In the Next-to-MSSM (NMSSM)~\cite{Ellwanger:2009dp},
a $Z_3$ symmetry is imposed under which all chiral MSSM fields including $S$ has a charge $1$.
Then the cubic term, $W=\kappa S^3$, is also allowed and $S$ can obtain a VEV of desired order.
Another scenario is to impose a discrete $R$-symmetry such as $Z_{5R}$ or $Z_{7R}$
in order to generate a tadpole term, $W = \xi_F S$~\cite{Panagiotakopoulos:1998yw,Panagiotakopoulos:2000wp}.
Such a model is called nMSSM.\footnote{
	The term (\ref{SHH}) along with the MSSM Yukawa couplings preserves 
	Peccei-Quinn (PQ) symmetry, denoted by U(1)$_{\rm PQ}$, under which the fields rotate as
	$S \to e^{2i\theta} S, H_u \to e^{-i\theta}H_u, H_d\to e^{-i\theta}H_d$.
	In order to avoid the appearance of extremely light Goldstone boson after symmetry breaking of weak scale,
	we need to add terms which explicitly break U(1)$_{\rm PQ}$.
	The cubic term or tadpole term break the U(1)$_{\rm PQ}$.
}
There are also other choices of discrete $R$-symmetries which lead to more general superpotential of $S$~\cite{Lee:2011dya}.
It is known that the Higgs mass bound is relaxed in the singlet extensions such as NMSSM or nMSSM 
because the coupling (\ref{SHH}) induces a quartic coupling of the Higgs boson at tree-level,
which, in turn, contributes to the Higgs mass.
A limitation of this possibility is that the coupling constant $\lambda$ cannot be arbitrary large
because renormalization group flow makes $\lambda$ larger at higher energy scale,
and it may blow up before reaching to the grand unification theory (GUT) scale.
Also, the tree level contribution becomes small as $\tan\b$ is increased.
Thus $\tan\beta$ must be small in order to make the Higgs heavy.

The smallness of $\tan\beta$ is not preferred in the light of the muon anomalous magnetic moment (muon $g-2$). The latest studies reported that there exists a deviation between the measured value of the muon $g-2$~\cite{Bennett:2006fi} and the Standard Model prediction, at more than $3\sigma$ level~\cite{g-2_hagiwara, g-2_davier}. As pointed out in Ref.~\cite{Moroi:1995yh}, the deviation can be naturally explained in low-energy SUSY standard model. The leading contribution from SUSY particles to the muon $g-2$ is given by,
\begin{equation}
\Delta a_\mu \sim \frac{g_2^2}{32\pi^2}\frac{m_{\mu}^2}{m_{\rm SUSY}^2}\tan\beta,
\end{equation}
where $m_{\mu}$ and $m_{\rm SUSY}$ denote the muon mass and the mass of SUSY particles respectively. The contribution is proportional to $\tan\beta$. Thus if we take the deviation of the muon $g-2$ seriously, we need a scheme to explain the large Higgs mass with large $\tan\beta$, unless sleptons, the wino and/or the bino are very light.
Therefore, if we want to explain the LHC result for the Higgs mass and the muon $g-2$ simultaneously, 
we need both large $\lambda$ and $\tan\beta$, which, however, violates the perturbativity of $\lambda$.\footnote{There is another possibility. By adding vector-like matters to MSSM, the heavy Higgs mass as 130 GeV and the muon g-2 can by simultaneously explained with large $\tan\beta$~\cite{Endo:2011mc}.}

In this paper we show that it is possible to modify the running of the $\lambda$
so that the perturbativity bound is significantly relaxed.
Thus the coupling constant $\lambda$ can take much larger value at weak scale than usually thought.
Interestingly, we show that more drastic effects on the Higgs mass appears for such a large value of $\lambda$
through the radiative correction involving the neutralino and Higgs bosons,
which dominates over the usual radiative correction induced by the top Yukawa,
and this effect is significant for wide range of $\tan\beta$.
This is contrasted to the case of tree-level correction to the Higgs mass, 
which is efficient only for small $\tan\beta$.
The radiative correction in the NMSSM was calculated in the literature~\cite{nmssm1loop,Ellwanger:2005fh}, and in particular the effect of $\lambda$ 
in the loop corrections was
discussed in \cite{Ellwanger:2005fh}.
However, that effect was considered to be subdominant because of the naive perturbativity bound on $\lambda$.
We show that
the lightest Higgs can be as heavy as $130-140$\,GeV while keeping the perturbativity up to the GUT scale, even without the large trilinear coupling of stops in this setup.

In Sec.~\ref{sec:sing} we describe phenomenological aspects of nMSSM with large $\lambda$
and show that the Higgs mass can be sufficiently heavy.
In Sec.~\ref{sec:UV} a UV model is constructed which 
allows large $\lambda$ at low energy.
We summarize the results in Sec.~\ref{sec:sum}.

\section{Higgs mass bound in singlet extension of the MSSM}  \label{sec:sing}

Let us study the bound on the Higgs mass in singlet extension of the MSSM.
To be concrete, we focus on a particular setup, although detailed structures of
the model is not so important.

Although the NMSSM is an interesting scenario, as briefly explained in the Introduction,
it has some phenomenological problems.
The spontaneous breakdown of $Z_3$ symmetry 
predicts the cosmological domain wall formation.
Introducing a small $Z_3$ breaking in order to destabilize the domain wall 
is not successful since it reintroduces the hierarchy problem~\cite{Abel:1995wk}.\footnote{
	See, however, Ref.~\cite{Hamaguchi:2011nm}.
}

Here we consider another type of a singlet extension model, so-called nMSSM, in which
a tadpole term $W= \xi_F S$ exists.
The origin of the tadpole term and the magnitude of its coefficient is explained by 
some mechanisms~\cite{Panagiotakopoulos:1998yw,Panagiotakopoulos:2000wp,Ellwanger:2008py}. 
This model does not suffer from the domain wall problem and it may be a good starting point to consider
the phenomenology of a singlet extension model.
Concerning the lightest Higgs mass, a situation is similar to the NMSSM, but there is a little difference.
In the NMSSM, it is not so easy to obtain a large value for $\kappa$, the coefficient of the cubic term of $S$
in the superpotential,
although the model described in the next section allows the existence of this coupling. 
In that case, the parameter space in which we can increase the lightest Higgs boson mass may be small.
Such an additional coupling does not exist in the nMSSM and it is more easy to obtain a large lightest Higgs boson mass.

Let us consider the nMSSM as the low-energy effective theory,
although the existence of cubic term $\sim \kappa S^3$ does not much affect the following discussion
as long as $\kappa$ is small. The superpotential is given by
\begin{equation}
	W = y^u_{ij} Q_iU_j^cH_u + y_{ij}^d Q_iD_j^c H_d + y_{ij}^e L_iE_j^c H_d
	+ \lambda S H_u H_d + \xi_F S,
\end{equation}
where $i,j=\{1,2,3\}$ is the generation indices
and we take $\lambda$ and $\xi_F$ to be real by the field redefinition of $S$ and $H_u, H_d$.
For convenience, we take $\lambda > 0$ and $\xi_F > 0$ hereafter.
The scalar potential for the Higgs sector is given by
\begin{equation}
	V=V_F + V_D + V_{\rm soft},
\end{equation}
where
\begin{equation}
	V_F = \lambda^2 |S|^2(|H_u|^2+|H_d|^2) + |\lambda H_u H_d+ \xi_F|^2,
\end{equation}
\begin{equation}
	V_D = \frac{g^2+g'^2}{8}(|H_u|^2-|H_d|^2)^2 + \frac{g^2}{2}|H_d^\dagger H_u|^2,
\end{equation}
\begin{equation}
	V_{\rm soft} = m_{H_u}^2|H_u|^2 + m_{H_d}^2|H_d|^2 + m_S^2 |S|^2
	+(\xi_S S + {\rm h.c.}) + (A_\lambda \lambda S H_uH_d +{\rm h.c.}).
\end{equation}
Here $g$ and $g'$ are SU(2)$_L$ and U(1)$_Y$ gauge coupling constants, respectively.
$V_{\rm soft}$ comes from the SUSY breaking effect. 
In the following we assume that SUSY breaking parameters $A_\lambda, \xi_S$ are real and there are no 
sources of explicit CP violation, for simplicity. We also assume $m_S^2 > 0$.
A concrete scenario of SUSY breaking is not specified here, since the following discussion does not depend on
details of the SUSY breaking parameters.
We can set $H_d^-=0$ and $\langle H_d^0\rangle\equiv v_d (> 0)$ 
to be real by SU(2)$_L \times$ U(1)$_Y$ rotations.
From the condition $\partial V/\partial H_u^+ = 0$, we find that $H_u^+=0$ is only a minimum
if $m_{H_u}^2+\lambda^2 v_s^2+(g^2/2)v_d^2 + (g^2+g'^2)(|v_u|^2-v_d^2)/4 > 0$
~\cite{Menon:2004wv}.
Then, if $\xi_S < 0$ and $A_\lambda > 0 $, 
we find $v_S > 0$ and $v_u > 0$ to be also real
and there is no spontaneous CP violation~\cite{Romao:1986jy}. Explicitly,
\begin{equation}
	v_s = -\frac{\xi_S - \lambda A_\lambda v_u v_d}{m_S^2 + \lambda^2 v^2},
\end{equation}
where $v^2=v_u^2+v_d^2$.
In this model, the tree-level lightest Higgs mass is bounded as
\begin{equation}
	m_h^2 \leq m_Z^2 \cos ^2 (2\beta) + \lambda^2 v^2 \sin^2 (2\beta),  \label{tree}
\end{equation}
where the equality is saturated when the mixing between the singlet $S$ and Higgs boson is small.
The second term in (\ref{tree}) is a new contribution appearing in a singlet extension.
It is clear that the correction becomes important for a large value of $\lambda$.
There are studies on the upper bound on the lightest Higgs mass in such a singlet extension of the MSSM.\footnote{See, e.g., \cite{Barger:2006dh}
for a comparison of the Higgs masses in several singlet extension models.}
Any such effort suffers from the fact that $\lambda$ must be large in order to make the Higgs heavy
but this tends to break down the perturbativity of $\lambda$ itself below the GUT scale.
It is limited as $\lambda \lesssim 0.7$ at the weak scale and this restricts the effect on the Higgs mass,
although the existence of extra matters may relax the bound to some extent 
(see e.g., Refs.~\cite{Masip:1998jc,Barbieri:2007tu}).
Note that $\tan\beta$ cannot be large in order for the additional contribution to be significant,
but it is not favored if we want to explain the muon anomalous magnetic moment by SUSY contributions.

For larger $\lambda$ at the weak scale, the effect on the Higgs is more drastic.
First, apparently, the contribution to the tree-level mass (\ref{tree}) increases.
Second, more importantly, the radiative correction to the Higgs mass from the $\lambda$ coupling
starts to dominate over the usual top Yukawa contribution.
Therefore, the light Higgs mass increases for larger $\lambda$ more rapidly 
than the usual thought.
Moreover, the radiative correction is not suppressed by $1/\tan\beta$ as opposed to the
tree-level correction in (\ref{tree}), hence the Higgs mass can be large as $130-140$\,GeV, with large $\tan\beta$.
This is welcome from the viewpoint of muon anomalous magnetic moment.
In the next section we explicitly construct a UV model which allows a large value of $\lambda$
without disturbing the perturbativity.
For a while, we simply analyze the low-energy model with large $\lambda$.

To see how the Coleman-Weinberg (CW) effective potential~\cite{Coleman:1973jx}
affects the Higgs mass we must sum up contributions from neutralino and chargino loops
as well as neutral and charged Higgs boson loops~\cite{Ellwanger:2005fh}.
The CW potential is given by
\begin{equation}
	V_{\rm CW} = \frac{1}{64\pi^2}{\rm STr}\mathcal M^4 \left[ \log \frac{\mathcal M^2}{Q^2} -\frac{3}{2}	
	\right].
\end{equation}
The bosonic part includes the up and down type Higgs and the singlet scalar, whose mass matrix is given by
\begin{equation}
	\mathcal M_S^2 = \begin{pmatrix}
		m_{H_u}^2+\lambda^2(v_s^2+v_d^2)
		&& -\lambda \xi_F -  \lambda A_\lambda v_s+ 2\lambda^2 v_uv_d 
		&& 2\lambda^2 v_s v_u - \lambda A_\lambda v_d  \\
		~   &&   m_{H_d}^2+\lambda^2(v_s^2+v_u^2)
		&& 2\lambda^2 v_s v_d - \lambda A_\lambda v_u \\
		~   &&    ~    &&   m_S^2+ \lambda^2 v^2
	\end{pmatrix},
\end{equation}
for the CP-even scalars, and 
\begin{equation}
	\mathcal M_P^2 = \begin{pmatrix}
		m_{H_u}^2+\lambda^2(v_s^2+v_d^2)
		&& \lambda \xi_F + \lambda A_\lambda v_s   
		&& \lambda A_\lambda v_d  \\
		~   &&   m_{H_d}^2+\lambda^2(v_s^2+v_u^2)
		&& \lambda A_\lambda v_u \\
		~   &&    ~    &&   m_S^2+ \lambda^2 v^2
	\end{pmatrix},
\end{equation}
for the CP-odd scalars. The singlino and neutral higgsino squared mass matrix is given by\footnote{
	In these expressions we have ignored the gauge coupling dependent terms, i.e.,
	we took the limit of small $g$ and $g'$, since we are interested in the dominant 
	$\lambda$ correction.
	In this limit, mixing between singlino, higgsino and gauginos are dropped.
	We have checked that the effects of the mixing are rather small and can be neglected.
}
\begin{equation}
	\mathcal M_F^2 = \begin{pmatrix}
		\lambda^2(v_s^2+v_d^2)   &&   \lambda^2 v_uv_d   && \lambda^2 v_s v_u   \\
		~   &&   \lambda^2( v_s^2 + v_u^2 )  && \lambda^2  v_s v_d \\
		~   &&    ~    &&   \lambda^2 v^2
	\end{pmatrix}.
\end{equation}
In the limit of large $\tan\b$, small $A_\l$ and large scalar masses, the leading contribution
to the lightest Higgs mass in the presence of SUSY breaking takes the form of~\footnote{In numerical calculations, we used the complete form of $V_{\rm CW}$.}
\begin{eqnarray}
\Delta m_h^2 &\simeq & \frac{\sin^2\beta}{2}\left(\frac{\partial^2}{\partial v_u^2} - \frac{1}{v_u}\frac{\partial}{\partial v_u}\right) V_{\rm CW} , \nonumber \\
&\simeq& \frac{\sin^2\beta}{4\pi^2} \lambda^4 v_u^2 \log\frac{M_s^2}{M_f^2} , \label{eq:mhcor}
\end{eqnarray}
where $M_s^2$ and $M_f^2$ are the typical value of scalar masses, $M_s^2 \sim m_S m_{H_d}$ and fermion masses, $M_f^2 \sim \lambda^2 (v_s^2 + v_u^2)$, respectively.
This correction may exceed the contribution from the usual one from top and stop loops for $\lambda \gtrsim 1$.

\begin{figure}[tbp]
\begin{center}
\includegraphics[scale=1.0]{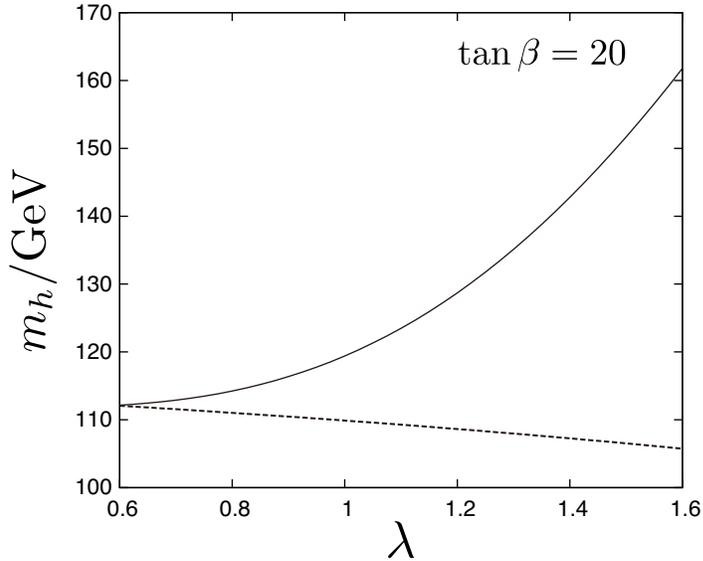}
\caption{
	The lightest Higgs mass as a function of $\lambda$ for $\tan\beta=20$.
	The upper line includes the radiative correction from neutralino and Higgs loops
	and the lower line does not. The other parametes are set to be $\lambda v_S = 200\, {\rm GeV}$, $A_{\lambda} = 0\,  {\rm GeV}$, $\xi_F=(200\, {\rm GeV})^2$ and $\xi_S = -(1000\,  {\rm GeV})^3$. The soft masses $m_{H_u}^2$, $m_{H_d}^2$ and $m_S^2$ are chosen so that the electroweak symmetry breaks successfully. The soft masses and the trilinear coupling of stops are taken to be 1.5 TeV and 0 GeV, respectively.
}
\label{fig:mh}
\end{center}
\end{figure}

\begin{figure}[tbp]
\begin{center}
\includegraphics[scale=1.0]{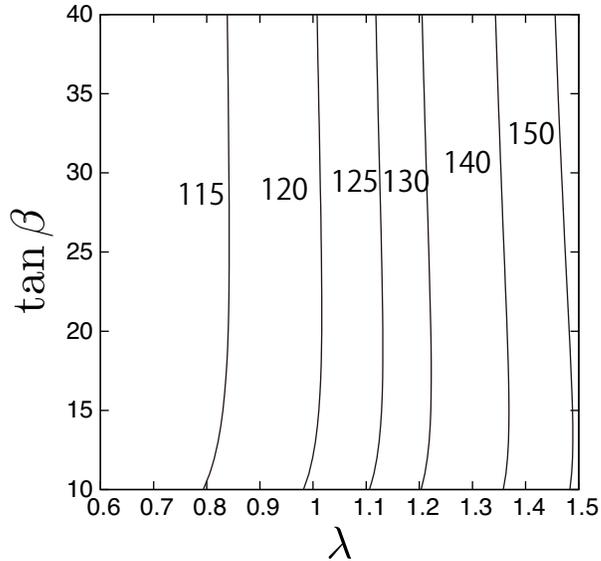}
\caption{
The contours of the lightest Higgs mass are shown in the unit of {\rm GeV}, on $\lambda$ vs. $\tan\beta$ plane. The parameters are same as Fig.~{\ref{fig:mh}}.
 }
\label{fig:mh_tb}
\end{center}
\end{figure}

We have calculated Higgs mass by including the radiative corrections from the neutralinos and the neutral Higgs as well as the top/stop corrections.
Fig.~\ref{fig:mh} shows the Higgs mass as a function of $\lambda$.
The upper solid line includes the radiative correction from neutralino and Higgs loops
and the lower dashed line does not. The Heavy scalar masses are about 1 {\rm TeV} and 2 {\rm TeV} for $H_d$ like and $S$ like state, respectively, with the chosen parameters.
It is seen that the Higgs mass, $m_h$, can be as large as 130\,GeV for $\lambda \simeq 1.2$.
This is clearly due to the large radiative correction explained above. 
In Fig.~\ref{fig:mh_tb}, we also show contours of the lightest Higgs mass 
on $\lambda$-$\tan\beta$ plane. One can clearly see that
 the mass of the lightest Higgs is insensitive to $\tan\beta$. 
Thus, we can easily explain the deviation of muon $g-2$, consistently with the Higgs mass larger than $130{\rm GeV}$. With such a large value of $\lambda$, $\lambda \simeq 1.2$, it is difficult to keep the perturbativity up to the GUT scale. However, as will be shown in the next section, in the setup of our UV model, even this large value of $\lambda$ does not lead to the breakdown of perturbativity up to the GUT scale.

%
%
%
Let us briefly comment on the little hierarchy problem. In our model, the lightest Higgs mass can be raised by the radiative corrections from neutralinos and neutral Higgs, without relying on those from top/stops. Therefore stops can be fairly light as $\mathcal{O}(100)$\,GeV,\footnote{The recent constraints from SUSY search can be avoided, e.g., when the SUSY particle masses are degenerated.} so that radiative corrections to $m_{H_u}^2$ can be of order of $\mathcal{O}(10^4-10^5)\,{\rm GeV}^2$.
 On the other hand, the radiative corrections from Higgs are also important since $\lambda$ is large. It is approximately given by,
\begin{eqnarray}
\delta m_{H_u}^2 \simeq \frac{\lambda^2}{8\pi^2} ( m_{H_u}^2 + m_{H_d}^2 + m_S^2 + |A_\lambda|^2 ) \ln \frac{M_s}{M_{\rm Mess}},
\end{eqnarray}
where $M_{\rm Mess}$ is the messenger scale. The correction to $m_{H_u}^2$ is portional to $\lambda^2$, while the corrections to the lightest CP-even Higgs mass is proportional to $\lambda^4$ (see. Eq.(\ref{eq:mhcor})). Therefore when we take larger $\lambda$ and small values of $m_{H_d}^2$ and $m_S^2$, e.g., $\mathcal{O}(10^5)\,{\rm GeV}^2$, the fine-tuning among the tree level values, $(m_{H_u}^2)_{\rm tree}$, the squared of the $\mu$ term and $\delta m_{H_u}^2$, is reduced, compared to the case that the mass of the lightest Higgs is raised by top/stop loops.

\section{UV completion}   \label{sec:UV}

In this section we construct a UV complete model which allows a large $\lambda$ in any singlet extension of the MSSM.
For this purpose, we use strong dynamics at high energy scale. 
The fat Higgs models~\cite{Harnik:2003rs} also achieve a large $\lambda$ by some strong dynamics, but 
there is a crucial difference of our model from the fat Higgs models. In our model, all the fields
(including the singlet and the Higgs fields) are elementary fields rather than composites.
This makes the calculation more reliable, and also, arbitrary superpotential of the singlet field can be introduced.
The perturbative gauge coupling unification is perfectly maintained. There are also models based on warped five dimensions~\cite{Birkedal:2004zx},
which have some similarity to our model. It may be interesting to investigate the relation between the four dimensional and five dimensional models
in terms of the AdS/CFT duality~\cite{Maldacena:1997re}.

\subsection{Basic strategy}\label{sec:basic strategy}

The general singlet extension of the MSSM has the following superpotential
\beq
W= y_u H_u Q U^c+y_d H_d Q D^c+y_\ell H_d L E^c+\lambda S H_u H_d +W_S, \label{eq:nmssmsuperpot}
\eeq
where $W_S$ is an arbitrary superpotential for the singlet $S$. The renormalization group (RG) equation for $\lambda$ is
\beq
\frac{d \lambda}{d t}=(\g_S+\g_{H_u}+\g_{H_d})\l,
\eeq
where we denote the anomalous dimension of a chiral field $\Phi$ by $\g_\Phi$,
and $t$ is the log of the renormalization scale.
In all the calculable cases, the anomalous dimension of a gauge singlet field is positive, $\g_S>0$.
$\g_{H_u}$ and $\g_{H_d}$ get negative contributions from the $\SU(2)_L \times \U(1)_Y$ interactions,
but the couplings of these gauge groups are not so large. Then, the sum $\g_S+\g_{H_u}+\g_{H_d}$ is positive
and $\lambda$ becomes large as the energy scale is increased. If we demand that there is no Landau pole of $\l$ below the GUT scale,
$\l$ is bounded from above at the electroweak scale. This conclusion seems inevitable as long as $S$ is an elementally gauge singlet field.

However, we can obtain a large value for $\l$ even if $S$ is a singlet in the following way.
First, we introduce chiral matter fields, $G$ and ${\tilde G}$, 
which transform as the fundamental and anti-fundamental 
representations of a new hidden gauge group $\SU(5)_{\rm hid}$. Then we replace the above superpotential by
\beq
W \to y^{\rm UV}_u H^{\rm UV}_u Q U^c+y_d^{\rm UV} H_d^{\rm UV} Q D^c+y_\ell^{\rm UV} H_d^{\rm UV} L E^c+\lambda^{\rm UV} S {\tilde G} G +W_S+W_{\rm hid}, \label{eq:fullsuperpot}
\eeq
where $W_{\rm hid}$ will be described later. The superscript ``UV'' in the yukawa couplings and the Higgs fields means that 
these are the ones in the UV theory, which will be different from the low energy ones as described below.

Next, let us suppose that the gauge group $G^{\rm UV}_{\rm SM} \times \SU(5)_{\rm hid}$ (where 
$G^{\rm UV}_{\rm SM}=\SU(3)_{C} \times \SU(2)_{L} \times \U(1)_{Y}$ is the SM gauge group in the UV) is broken
down to a subgroup 
\beq
G^{\rm UV}_{\rm SM} \times \SU(5)_{\rm hid} \to G_{\rm SM}.
\eeq
This symmetry breaking can be achieved by the VEV of some chiral field $B$ which is in the bifundamental representation of
$\SU(5)^{\rm UV}_{\rm GUT} \times \SU(5)_{\rm hid}$, where $\SU(5)^{\rm UV}_{\rm GUT} (\supset G^{\rm UV}_{\rm SM})$ is the
usual $\SU(5)$ GUT gauge group. In this case, the gauge couplings $g_a~(a=1,2,3)$ of the SM gauge group $G_{\rm SM}$
are related to the couplings $g^{\rm UV}_a$ and $g_{\rm hid}$ of $G^{\rm UV}_{\rm SM} \times \SU(5)_{\rm hid}$ as
\beq
\frac{1}{(g_a)^2}=\frac{1}{(g^{\rm UV}_a)^2}+\frac{1}{(g_{\rm hid})^2}~~~(a=1,2,3). \label{eq:couplingmatching}
\eeq
This change of couplings does not spoil the gauge coupling unification, at least at the one loop level.

Below the symmetry breaking scale, $G$ and $\tilde{G}$ transform as the fundamental and anti-fundamental
representations of $\SU(5)_{\rm GUT}$.
Let us further suppose that $H_u^{\rm UV}$ ($H_d^{\rm UV}$) mixes with the $\SU(2)_L$ doublet component $G_2$ of $G$ ($\tilde{G}_2$ of $\tilde{G}$)
as (see also Ref.~\cite{Birkedal:2004zx})
\beq
\left(
\begin{array}{c}
H_u^{\rm UV} \\
G_2
\end{array}
\right)
&=&
\left(
\begin{array}{cc}
\cos \h_1 & -\sin \h_1 \\
\sin \h_1 & \cos \h_1 \\
\end{array}
\right)
\left(
\begin{array}{c}
H_u \\
G'_2
\end{array}
\right), \nonumber \\
\label{eq:mixingGH}\\
\left(
\begin{array}{c}
H_d^{\rm UV} \\
\tilde{G}_2
\end{array}
\right)
&=&
\left(
\begin{array}{cc}
\cos \h_2 & -\sin \h_2 \\
\sin \h_2 & \cos \h_2 \\
\end{array}
\right)
\left(
\begin{array}{c}
H_d \\
\tilde{G}'_2
\end{array}
\right). \nonumber
\eeq
If $G$ and $\tilde{G}$ have supersymmetric mass terms, these fields can be integrated out.
Then, at low energies, we get back to the superpotential (\ref{eq:nmssmsuperpot}) with
\beq
y_u=y_u^{\rm UV} \cos\h_1,~~~y_d=y_d^{\rm UV} \cos\h_2,~~~y_\ell=y_\ell^{\rm UV} \cos\h_2,~~~\l=\l^{\rm UV} \sin\h_1 \sin\h_2 . \label{eq:coulplingmatching2}
\eeq

The important point is that the RG equation for $\l^{\rm UV}$ is now given by
\beq
\frac{d \l^{\rm UV}}{d t}=(\g_S+\g_{G}+\g_{\tilde{G}})\l^{\rm UV},
\eeq
and $\g_{G},\g_{\tilde G}$ receive negative contributions from the gauge group $\SU(5)_{\rm hid}$.
Thus, if $\SU(5)_{\rm hid}$ is strongly coupled, the sum $\g_S+\g_{G}+\g_{\tilde{G}}$ can be negative and 
$\l^{\rm UV}$ becomes larger as we flow to low energies. Therefore, we can obtain a large value for $\l$ at the electroweak scale
if the mixing parameters $\sin\h_1$ and $\sin\h_2$ are not too small.

\subsection{The model}

\begin{table}[Ht]
 \begin{center}
 \begin{tabular}{|c|c|c|}
 \hline 
&$\SU(5)_{\rm GUT}^{\rm UV}$&$\U(5)_{\rm hid}=\SU(5)_{\rm hid}+\U(1)_{\rm hid}$\\ \hline
$G~~(\tilde{G})$& ${\bf 1}$ & ${\bf 5}~~(\bar{\bf 5})$ \\ \hline
$F~~(\tilde{F})$& ${\bf 1}$ & ${\bf 5}~~(\bar{\bf 5})$ \\ \hline
$B~~(\tilde{B})$&$\bar{\bf 5}~~({\bf 5})$ & ${\bf 5}~~(\bar{\bf 5})$  \\ \hline
$\Phi=\Phi_S+\Phi_A$& ${\bf 1}$ &${\bf 5} \times \bar{\bf 5}={\bf 1+{\rm Adj}}$\\ \hline
\end{tabular}
\caption{The matter content of the hidden sector model.}
\label{table:1} 
\end{center}
\end{table}

The matter content of the model is shown in Table~\ref{table:1}. We have introduced the $\U(1)_{\rm hid}$ gauge group in addition to the 
$\SU(5)_{\rm hid}$. The role of $\U(1)_{\rm hid}$ is just to give a mass to some unwanted moduli field by the Higgs mechanism. 
The superpotential $W_{\rm hid}$ in Eq.~(\ref{eq:fullsuperpot}) consists of two parts,
\beq
W_{\rm hid}=W_1+W_2.
\eeq
$W_1$ and $W_2$ are given by
\beq
W_1= \xi \tr\left[ (\Phi_S+\eta \Phi_A)B\tilde{B} - \mu^2 \Phi_S\right],
\eeq
and 
\beq
W_2= \tilde{F}(m_1G+\eta_1 B H)+F(m_2 \tilde{G}+\eta_2 \tilde{B} \tilde{H}).
\eeq
where we have defined
\beq
H=(0,H_u^{\rm UV}),~~~\tilde{H}=(0,H_d^{\rm UV}).
\eeq
In the following, we pretend as if $H,\tilde{H}$ are complete multiplets of $\SU(5)_{\rm GUT}$.

The superpotential $W_1$ is introduced to give VEVs to the bifundamental fields $B$ and $\tilde{B}$.
The equations of motion of $\Phi$ and the D-potential of the gauge groups
set $\vev{B}=\langle {\tilde{B}} \rangle=\mu \cdot {\mathbf 1}$, up to gauge transformation.
Then, $\SU(5)_{\rm GUT}^{\rm UV} \times \U(5)_{\rm hid}$ is broken down to $\SU(5)_{\rm GUT}$,
and the superpotential $W_2$ becomes
\beq
W_2 \to \tilde{F}(m_1G+\eta_1 \mu H)+F(m_2 \tilde{G}+\eta_2 \mu \tilde{H}).
\eeq
From this superpotential, $G' \propto m_1G+\eta_1 \mu H$ and $\tilde{G}' \propto m_2 \tilde{G}+\eta_2 \mu \tilde{H}$ obtain Dirac masses with
the fields $\tilde{F}$ and $F$. Then, there remain the low energy Higgs fields $H_u$ and $H_d$ given in Eq.~(\ref{eq:mixingGH}) with the mixing
angle $\tan \theta_i=-\eta_i \mu/m_i~~(i=1,2)$. We assume that $\eta_i \mu$ and $m_i$ are of the same order.\footnote{It may be easy to generate
$\mu$ and $m_i$ of the same order by replacing them with some chiral field which dynamically develops a VEV. In this paper we treat them just
as free parameters of the model.}

Let us investigate the anomalous dimensions of the fields.
For our mechanism to work, the anomalous dimension of $G$ and $\tilde{G}$, $\g_G$, must be large and negative.
Large negative value of $\g_G$ is achieved by the dynamics of $\SU(5)_{\rm hid}$. 
In the present model, $\SU(5)_{\rm hid}$ has conformal fixed points. The details of the fixed point depend
on which operators in the superpotential are on the fixed point.
We consider the following two cases.
\begin{itemize}
\item Case 1 : The operator $\xi\eta \tr \Phi_A B\tilde{B}$ is on the fixed point.
\item Case 2 : The operators $\xi\eta \tr \Phi_A B\tilde{B}$ and $\l^{\rm UV}SG\tilde{G}$ are on the fixed point.
\end{itemize}
Couplings other than the ones above are assumed to be small so as not to disturb the fixed point dynamics.
The reason that we take $\tr \Phi_A B\tilde{B}$ to be on the fixed point in both cases will be explained below.

The anomalous dimensions can be determined exactly at the conformal fixed point even in strongly coupled theories,
by using the $a$-maximization technique of Ref.~\cite{Intriligator:2003jj}.
The result is listed in Table~\ref{table:2}.
In the case 1, the large negative value of $\g_G$ makes $\lambda^{\rm UV}$ decreasing as the energy scale is increased, as explained in the previous subsection. 
In the case 2, the coupling $\l^{\rm UV}$ itself is on the conformal fixed point, and is constant at high energies.
In fact, $\g_S+2\g_G=0$ at the conformal fixed point.
If we make a very crude ``approximation'' that $\g_S$ is given by the one-loop expression
\beq
\g_S \sim \frac{5}{16\pi^2}(\l^{\rm UV})^2,
\eeq
then by using the value $\g_S=0.705$, we obtain the fixed point value of $\l^{\rm UV}$,
\beq
\l^{\rm UV} \sim 4.7~~.\label{eq:roughlambda}
\eeq
In the next subsection, we will see the explicit RG flows of the couplings.

\begin{table}[Ht]
\begin{center}
\begin{tabular}{|c|c|c|c|c|c|c|}
\hline
&$\g_G$&$\g_F$&$\g_B$&$\g_{\Phi_A}$&$\g_{\Phi_S}$&$\g_S$ \\ \hline
case 1 &-0.371&-0.371&0.151&-0.303&0&0 \\ \hline
case 2 &-0.352&-0.376&0.154&-0.308&0&0.705\\ \hline
\end{tabular}
\caption{The anomalous dimensions at the conformal fixed points. Contributions from couplings not relevant to the fixed points are neglected.
At the fixed points, $\g_{\tilde G}=\g_G$, $\g_{\tilde F}=\g_F$ and $\g_{\tilde B}=\g_B$.}
\label{table:2} 
\end{center}
\end{table}

Finally, let us explain the reason why we take the operator $\tr\Phi_A B\tilde{B}$ on the conformal fixed point~\cite{Sato:2009yt}.
The fields $B, \tilde{B}$ are charged under the $\SU(5)_{\rm GUT}^{\rm UV}$ gauge group, so it contributes to the running
of the SM gauge couplings. Naively, $B$ and $\tilde{B}$ contribute to the $\beta$ functions of the SM gauge couplings as 5 flavors of ${\bf 5}+\bar{\bf 5}$
representations. Then we must be careful about the Landau pole of the SM gauge couplings, especially when the mass scale $\mu$ of the model is small.
However, this is not correct due to the strong dynamics at the conformal fixed point.
The exact $\beta$ function in SUSY theory is given in general by~\cite{Novikov:1983uc}\footnote{Notice that our convention for anomalous dimensions
differs from that of some of the literature by a factor of 2.}
\beq
\b(g)=\m \frac{\q}{\q \m} g^2=-\frac{g^4}{8\pi^2}\frac{3t(A)-\sum_{i} (1-2\g_i)t(i)}{1-t(A)g^2/8\pi^2},\label{eq:NSVZ}
\eeq
where $t(i)$ is the Dynkin index of the matter field labeled by $i$, $t(A)$ is the Dynkin index of the adjoint representation, and $\g_i$ is the anomalous dimension of the matter $i$. 
From this $\beta$ function, we can see that the contribution of $B$ and $\tilde{B}$ 
to the SM gauge coupling is not $5$, but
\beq
N_{\rm eff} \equiv 5(1-2\g_B). \label{eq:effectivenumber}
\eeq
Therefore, if $\g_B>0$, this contribution is effectively reduced. 
The anomalous dimension for a gauge non-singlet field at the conformal fixed point
is typically negative. However it can become positive when the coupling with the adjoint field, $\tr\Phi_A B\tilde{B}$, is present in the superpotential.
See Ref.~\cite{Sato:2009yt} where this mechanism is discussed in detail.\footnote{In fact, the fixed point of the case 1 is precisely the same as one of the fixed points discussed in Ref.~\cite{Sato:2009yt}.} In the present model, we have $N_{\rm eff} \simeq 3.5$ from the Table~\ref{table:2}.
Thus we need not worry about the Landau pole problem of the SM gauge coupling.

\subsection{RG flow}

The value of $\l$ at the weak scale is obtained by solving the RG equations. 
We solve them in the one-loop approximation.
Below the scale $\mu$, the RG equation is the same as the usual NMSSM (see e.g. Ref.~\cite{Ellwanger:2009dp}). 
At the scale $\mu$,
the couplings are changed as described in Eqs.~(\ref{eq:couplingmatching},\ref{eq:coulplingmatching2}). 
We approximate $g_a \simeq g_a^{\rm UV}$
in Eq.~(\ref{eq:couplingmatching}) since $g_{\rm hid}$ is very large. 

Above the scale $\mu$, the RG equations are given as follows. We neglect the MSSM yukawa couplings other than the top yukawa coupling $y_t$.
We omit the superscript ``UV'' of the couplings $y_u^{\rm UV}, y_d^{\rm UV}, y_\ell^{\rm UV}$ 
and $\l^{\rm UV}$ for simplicity.
The superpotential $W_S$ is taken as
\beq 
W_S=\xi_F S+\frac{1}{2}\mu' S^2+\frac{1}{3}\kappa S^3.
\eeq
Then, the high energy RG equations are given by
\beq
\frac{d g_a}{dt} &=& (b_a^{\rm MSSM}+N_{\rm eff})\frac{g^3_a}{16\pi^2}~~~(a=1,2,3), \\
\frac{d y_t}{dt} &=& \left( 6y_t^2-\frac{16}{3}g^2_3-3g^2_2 -\frac{13}{15}g^2_1\right)\frac{y_t}{16\pi^2}, \\
\frac{d \l}{dt} &=&  (\g_S+2\g_G) \l, \\
\frac{d \kappa}{dt} &=& 3\g_S \kappa.
\eeq
where $b_a^{\rm MSSM}$ is the usual MSSM contributions to the $\beta$ functions, and $N_{\rm eff}$ is defined in Eq.~(\ref{eq:effectivenumber}).
In the case 1, $\l$ (and $\kappa$) are assumed to be small. Then we have
\beq
\g_S=\frac{1}{16\pi^2}(5\l^2+2\k^2).
\eeq
In the case 2, $\g_S$ is the one given in Table~\ref{table:2}. In both cases, $\g_G$ is given Table~\ref{table:2}.

\begin{figure}
\begin{center}
\includegraphics[scale=0.9]{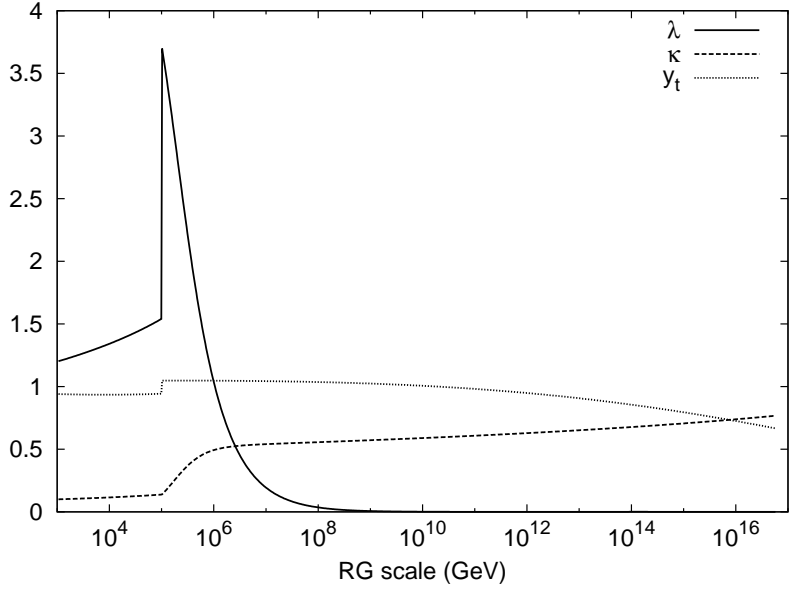}
\caption{An example of the RG flow in the case 1. The parameters are taken as $\cos\theta_1=0.9,~\cos\theta_2=0.3$ and $\mu =10^5~\GeV$.}
\label{fig:3-1}
\end{center}
\end{figure}

\begin{figure}
\begin{center}
\includegraphics[scale=0.9]{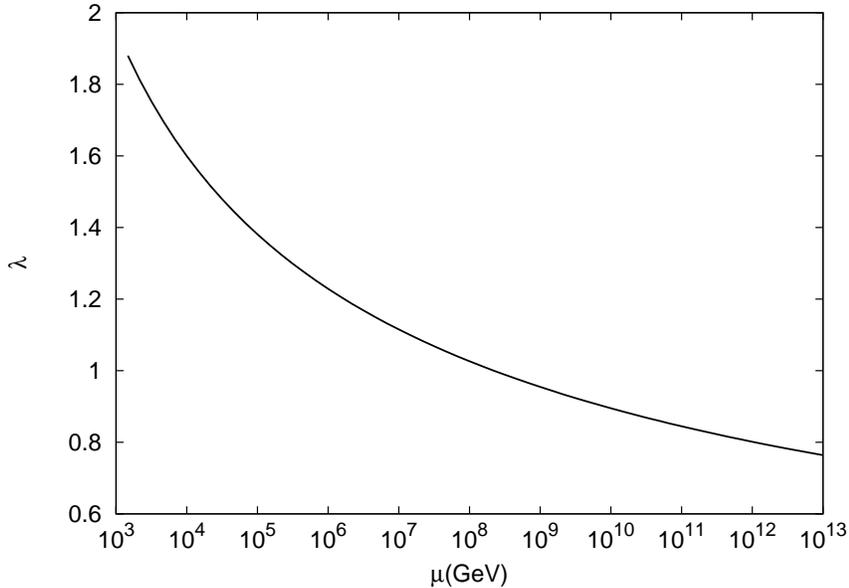}
\caption{The value of $\l$ at the weak scale as we change the scale $\mu$ in the case 2. The parameters are taken as $\cos\theta_1=0.9$ and
$\cos\theta_2=0.3$ and $\kappa=0$. We have used the value given in Eq.~(\ref{eq:roughlambda}) for the conformal fixed point value of $\lambda$.}
\label{fig:3-2}
\end{center}
\end{figure}

In Fig.~\ref{fig:3-1}, we give an example of the RG flow in the case 1.
In that example, parameters are taken as
\beq
\cos\theta_1=0.9,~~~\cos\theta_2=0.3,~~~\mu =10^5~\GeV. \label{eq:exampleparameter}
\eeq
Also, $\tan\beta$ is assumed to be large enough so that $y_t$ at the electroweak scale is almost the same as the SM value.
In the figure, the jumps in the couplings $\l$ and $y_t$ at the scale $\mu=10^5~\GeV$ is due to the change of the couplings 
given by Eqs.~(\ref{eq:couplingmatching},\ref{eq:coulplingmatching2}). In choosing the parameter $\cos\theta_1$, one have to be careful
because the Landau pole of the top yukawa $y_t$ may appear at high energies if $\cos\theta_1$ is too small\footnote{
The small value of the mixing angle, $\theta_2$ is also constrained by the Landou pole of the bottom Yukawa coupling, depending on $\tan\beta$.
}.
From the figure, we can clearly see the mechanism discussed in subsection~\ref{sec:basic strategy}; that is, $\l$ (or more precisely $\l^{\rm UV}$)
quickly decreases as the energy scale is raised above the scale $\mu$.

In Fig.~\ref{fig:3-2}, we give the value of $\l$ at the weak scale as we change the scale $\mu$ in the case 2.
The parameters $\cos\theta_1$ and $\cos\theta_2$ are taken the same as in Eq.~(\ref{eq:exampleparameter}),
and we take the conformal fixed point value of $\l$ as in Eq.~(\ref{eq:roughlambda}). 
If $\kappa$ is non-zero, it quickly blows-up due to the large value of $\gamma_S$, so we have set $\kappa=0$.
In the case 2, the calculation is not so reliable because of the large value of $\l$ near the threshold scale $\mu$. 
Therefore the result should be regarded only as a rough estimation. But our estimation may still be more reliable than the one in the fat Higgs models
because we need not use the so-called naive dimensional analysis at all.

\section{Summary}      \label{sec:sum}

In this paper we have studied how the upper bound on the lightest Higgs mass is relaxed
in a singlet extension of the MSSM with coupling like $\lambda S H_u H_d$, even at a large $\tan\beta$ region, 
which is preferred to explain the measured value of the muon anomalous magnetic moment.
It has been known that the coupling $\lambda$ cannot take an arbitrary large value
since it blows up as the energy scale is increased due to the renormalization group evolution,
and this fact limits the applicability of the singlet extension for the purpose of raising the Higgs mass.

We have revisited the issue of theoretical upper bound on the Higgs mass in such a scenario, since the recent LHC data may indicate relatively heavy Higgs boson.
We have explicitly constructed a UV model that avoids the blow up of the coupling constant $\lambda$
and allows a large value of $\lambda$ at the weak scale.
We found that for such large $\lambda$ the radiative correction to the Higgs mass from
neutralino, and Higgs boson loops become significant and 
makes the dominant correction to the Higgs mass.
Actually the Higgs mass can be large as 130\,GeV for $\lambda \sim 1.2$ at the weak scale
without disturbing perturbativity to the GUT scale.
This large radiative contribution survives at large $\tan\beta$, as opposed to the tree-level correction
from $\lambda$ which is necessarily suppressed for large $\tan\beta$.
This is an appealing feature of the present model since relative large $\tan\beta$ is needed
in order to explain the muon anomalous magnetic moment through SUSY contributions.

\section*{Acknowledgements}
We would like to thank K.~Hamaguchi and especially M.~Endo for useful discussion.
This work is supported in part by Grant-in-Aid for Scientific research from
the Ministry of Education, Science, Sports, and Culture (MEXT), Japan, 
No.\ 21111006 (K.N.), No.\ 22244030 (K.N.) and by JSPS 
Research Fellowships for Young Scientists (N.Y. and K.Y.)
and by World Premier International Research Center Initiative, MEXT, Japan.


\end{document}